# AC-frequency switchable correlated transports in rare-earth perovskite nickelates

*Jikun Chen[1]\*, Haifan Li[1], Jiaou Wang[2], Xinyou Ke[3], Binghui Ge[4], Jinhao Chen[1], Hongliang Dong[5],*

*Yong Jiang[1], and Nuofu Chen[6]*

[1] School of Materials Science and Engineering, University of Science and Technology Beijing, Beijing 100083, China

[2] Beijing Synchrotron Radiation Facility, Institute of High Energy Physics, Chinese Academy of Sciences, Beijing 100049, China

[3] Department of Mechanical and Aerospace Engineering, Case Western Reserve University, Cleveland, Ohio 44106, United States

[4] Institute of Physical Science and Information Technology, Anhui University, 230601, Heifei, Anhui, China

[5] Center for High Pressure Science and Technology Advanced Research, Shanghai 201203, China

[6] School of Renewable Energy, North China Electric Power University, Beijing 102206, China

Correspondence: Prof. Jikun Chen (jikunchen@ustb.edu.cn).

**Abstract:**

Whilst electron correlations were previously recognized to trigger beyond conventional direct current (DC) electronic transportations (e.g. metal-to-insulator transitions, bad metal, thermistors), their respective influences to the alternation current (AC) transport are largely overlooked. Herein, we demonstrate active regulations in the electronic functionalities of $d$-band correlated rare-earth nickelate ($Re$NiO$_3$) thin films, by simply utilizing their electronic responses to AC-frequencies ($f_{AC}$). Assisted by temperature dependent near edge X-ray absorption fine structure analysis, we discovered positive temperature dependences in Coulomb viscosity of $Re$NiO$_3$ that moderates their AC impedance ($R'+iR''$). Distinguished crosslinking among $R'$-$f_{AC}$ measured in nearby temperatures is observed that differs to conventional oxides. It enables active adjustability in correlated transports of $Re$NiO$_3$, among NTCR-, $T_{\text{Delta}}$- and PTCR- thermistors, via $f_{AC}$ from the electronic perspective without varying materials or device structures. The $T_{\text{Delta}}$-$f_{AC}$ relationship can be further widely adjusted via $Re$ composition and interfacial strains. The AC-frequency sensitivity discovered in $Re$NiO$_3$ brings in a new freedom to regulating and switching the device working states beyond the present semiconductor technologies. It opens a new paradigm for enriching novel electronic applications catering automatic transmission or artificial intelligence in sensing temperatures and frequencies.

The *d*-band electron correlation within transitional oxides and their respective applications in electronic devices beyond semiconductors is a central piece in modern condensed matter physics. The Coulomb energy domination in electronic orbital configurations and transitions enrich distinguished electronic functionalities, e.g. hydrogenation-induced quantum phase transitions [1-5], metal to insulator transitions (MIT) [6-12], broad temperature range thermistor [13,14], and bad metal transport [15]. The rare-earth nickelates (*Re*NiO$_3$) is one representative *d*-band correlated perovskites that exhibits exceptional sensitive electronic structures to external stimulus, such as temperature [6,9], lattice distortion [6,8], charge polarization [7], and chemical atmosphere [1-5]. In particular, the recent discoveries of the hydrogen induced electron localization within *Re*NiO$_3$ enables new electronic and Mottronic applications in ocean current sensing [1], synaptic plasticity [5], logic/memory devices [4], bio-sensing [2], and energy conversions [3].

Nevertheless, the previously achieved electronic and Mottronic functionalities within *d*-band correlated semiconductors (e.g. *Re*NiO$_3$) mainly rely on electronic responses to direct current (DC) signals [1-15], in which situation as-achieved functionalities are passively determined by material properties. In contrast, how to utilize their respective alternative current (AC) electronic transportations in correlated electronic devices remains yet unexplored. For instance, the insulating orbital configuration for *Re*NiO$_3$ (Ni$^{3+}$$t_{2g}^6 e_g^{1+\Delta}$ + Ni$^{3+}$$t_{2g}^6 e_g^{1-\Delta}$) was recognized to gradually transits with temperature [6,13], while more abrupt orbital transitions towards the metallic phase (Ni$^{3+}$$t_{2g}^6 e_g^1$) are triggered by elevating temperature (*T*) across a critical point (*T*$_{MIT}$) [6]. Whilst these orbital transitions beyond conventional semiconductors were previously recognized to enable DC-electronic properties such as thermistor and MIT [6,13], their respective regulations upon the AC-electronic functionalities remain as open questions. It is worthy to note that their Coulomb energy dominated orbital configuration will result in synergistic variations in their impedance (*R'*+i*R''*), as related to not only temperatures but also frequencies (*f*$_{AC}$). From this perspective, the previously demonstrated MIT and thermistor behaviors are barely the tip of the iceberg of the electronic functionalities within *Re*NiO$_3$, when their impedances are overwhelmed by DC-conduction (also the case for using low *f*$_{AC}$). With no doubt, exploring the beyond-conventional AC-transports at a higher *f*$_{AC}$ for correlated *Re*NiO$_3$ will open a new paradigm to discover new electronic functionalities beyond the present knowledge. More importantly, it also shed a light on utilizing *f*$_{AC}$ as an additional ***new freedom*** to achieve ***active regulation*** upon the working states of electronic devices from the electronic perspective without changing the material constitutions or device structures.

Herein, we demonstrate the distinguished *T* and *f*$_{AC}$ dependence in the AC transport for the correlated *Re*NiO$_3$ thin films, and accordingly enable actively switchable electronic functionalities as controlled by *f*$_{AC}$ without varying the materials and device structures. By simply increasing the input *f*$_{AC}$ of the detection AC signal, the electronic transportation behavior of *Re*NiO$_3$ transformed from a negative temperature coefficient of resistance (NTCR) thermistor at low *f*$_{AC}$, to a delta-temperature (*T*$_{Delta}$) thermistor at middle-range *f*$_{AC}$, and further towards a



positive temperature coefficient of resistance (PTCR) thermistor at high $f_{AC}$. Assisted by temperature dependent near edge X-ray absorption fine structure (NEXAFS) analysis, we demonstrate experimentally the temperature induced gradual orbital transitions within the insulating phase of $Re$NiO$_3$. From the perspective of materials design, we further show that the critical transition frequencies ($f_{NTCR-Delta}$ and $f_{Delta-PTCR}$) within a given range of temperature can be widely adjustable via the rare-earth composition ($Re$) or imparting interfacial strains.

The DC electronic conductivity ($\sigma_{DC}$) of $Re$NiO$_3$ as previously described by Drude model is written as: $\sigma_{DC} = \frac{nq^2\tau}{m^*}$, where $n$, $q$, $m^*$ represent for the concentration, charge and effective mass of the carrier, while $\tau$ is the life time between the carrier scatterings. Whilst the $Re$NiO$_3$ was previously demonstrated as a bad metal with a saturating $\tau$-$T$ tendency [15], the temperature dependence in $\sigma_{DC}$ originates from the $n$-$T$ and $m^*$-$T$ tendencies. Thermally activation of carriers is well recognized, as the concentration of semiconductors follows: $n_{(T)} = n_0 \exp\frac{E_g}{k_B T}$, where $k_B$ is the Boltzmann constant, $E_g$ is the energy band gap and $n_0$ is a constant initial carrier concentration [13,14]. Nevertheless, potential temperature dependency in $m^*$ was previously overlooked.

In contrast to conventional semiconductors, the carrier conduction within the insulating phase of $Re$NiO$_3$ is dominated by strong interactions with the lattice charge associated to the NiO$_6$ octahedron and follows a hopping mechanism [16]. As illustrated in Figure 1a, the carrier hopping among the adjacent NiO$_6$ octahedrons is expected to result in instantaneous charge polarizations, the electrical field from which drags the hopping carriers. This effect is analogical to increasing the viscosity of the carrier transport within semiconductors, e.g. by prohibiting the acceleration of carriers under externally imparted electrical field, and thereby enlarge $m^*$. Further elevating the temperature is expected to more significantly excite the carrier hopping together with the instantaneous polarized NiO$_6$, resulting in higher viscosity of the carrier transport. Following this consideration, potential $m^*$-$T$ tendencies within $Re$NiO$_3$ can be comprehended as a sign of temperature increased Coulomb viscosity ($\eta$) of the carrier transport, as defined herein to be $m^*_{(T)} = \eta_{(T)} m^*_0$.

From the perspective of DC-transport, a positive $\eta_{(T)}$-$T$ tendency moderates the exponential $n_{(T)}$-$T$ tendency, and results in more steady reduction in $R'$ at a broader range of temperature. Thus, it causes smaller down-shifting of $R'$-$f_{AC}$ curve when elevating the temperature, e.g. by per Kelvin. Meanwhile, the enlarged $\eta$ strengthens the electronic shielding of the lattice polarizations and thereby increases the relaxation frequency ($f_0$) from the perspective of AC-transport [17-24]. Thus, it causes larger right-shifting of $R'$-$f_{AC}$ curve when elevating the temperature, e.g. by per Kelvin. Combining the above two aspects, it may result in crosslinking among the $R'$-$f$ curves measured at nearby ranges of temperatures for the insulating phase of $Re$NiO$_3$. Hence, probing $R'$-$T$ using AC-source with selected $f_{AC}$ associated to the crosslinking region within the target range of temperature establishes a delta-shaped $R'$-$T$, which we nominated as $T_{Delta}$ thermistor transportation. Further enlarging or reducing $f$ beyond



the crosslinking region enables NTCR or PTCR thermistor transportations within *Re*NiO$_3$.

To realize the above concept, we epitaxially grew *Re*NiO$_3$ thin films, with compositions including SmNiO$_3$, NdNiO$_3$, EuNiO$_3$, HoNiO$_3$, and Sm$_{3/4}$Nd$_{1/4}$NiO$_3$ on single crystalline LaAlO$_3$ and SrTiO$_3$ substrates, using the solution based chemical spin coating followed by high oxygen pressure annealing according to our previous reports [13,14]. More experimental details are described in Supporting Information (Section A). Figure 1b demonstrates the cross section morphology of as-grown SmNiO$_3$/LaAlO$_3$ (001), where a coherent interface between the film and substrate is observed. The SmNiO$_3$/LaAlO$_3$ sample exhibit sharp MIT behavior with $T_{MIT}$ of ~120 ℃, as demonstrated in Figure 1c. To characterize the temperature induced variations in the electronic structure of as-grown SmNiO$_3$/LaAlO$_3$ (001), we performed temperature dependent NEXAFS to probe its Ni-*L* and O-*K* edges at various temperatures, as their results shown in Figure 1d and 1e, respectively. Reducing the temperature decreases the relative intensity in the proportion of sub-peak (B) within the Ni-$L_3$ spectrum (see Figure 1d), while the pre-peak (A) in O-*K* edge (from the $d^8L$ configuration) is also decreased (see Figure 1d). These observations indicate that the elevation in the insulating $t^6_{2g}e^2_g$ (Ni$^{2+}$) ground state orbital configuration compared to metallic $t^6_{2g}e^2_g$ (Ni$^{2+}$) $t^6_{2g}e^1_g$ (Ni$^{3+}$) [25,26] via descending temperature. Similar observations in NEXAFS spectrum were previously observed when strengthen the insulating phase of *Re*NiO$_3$ via reducing the size of *Re* [13] or imparting tensile lattice distortions [27]. It is also interesting to note the rising of an additional sub-peak (C) rise prior to the main peak (A) in the Ni-$L_3$ spectrum at low temperatures, indicating the dragging of empty orbitals partially towards a lower energy level. This observation is in consistency to the situation when the conduction band edge caves in downwardly that increases the band curvature and reduces the effective mass, resulting in a positive $\eta$-*T* tendency.

The impedances for SmNiO$_3$/LaAlO$_3$ (001) measure at various temperatures are plotted as a function of the AC frequency in Figure 2a and 2b for *R*' and *R*'', respectively. The AC-transport of the quasi-single crystalline SmNiO$_3$/LaAlO$_3$ is well described by equivalent circuit composed of a resistance and capacitor in parallel, as the respective core-core spectrums and fittings shown in Figure S1. At each temperature, *R*' and *R*'' follows the following equations [17]:

$$R' = \frac{\varepsilon_0 \sigma_{DC}}{\sigma_{DC}^2 C_0 + (2\pi f_{AC})^2 \varepsilon_0^2 \varepsilon'^2 C_0}$$

$$R'' = \frac{\varepsilon'}{2\pi f_{AC} c_0 \varepsilon'^2 + \sigma_{DC}^2 (2\pi f_{AC} C_0)^{-1}}$$

where $\varepsilon_0$ and $\varepsilon$' represent for the vacuum and relative permittivity, respectively, while $C_0$ is the geometric capacitance parameter in constant value. At low $f_{AC}$, the carrier transport is dominated by $\sigma_{DC}$ similar to DC-transport, in which situation *R*' shows a plateau magnitude of $R_{DC}=\varepsilon_0(\sigma_{DC}C_0)^{-1}$ while *R*'' is small. Increasing $f_{AC}$ weakens the carrier shielding of lattice polarization and enlarges *R*'' until reaching the maximum magnitude at its relaxation frequency $f_0$, at which point *R*' started to be reduced. By further varying the temperature, a negative $R'$-$T$ tendency is observed at low $f_{AC}$, while the respective $f_0$-$T$ tendency is positive. Thus, the $R'$-$f_{AC}$ curves measured at adjacently measured temperatures indeed cross-linked, as shown by the region marked in pink color in Figure 2a. Such pronounced cross-linking in $R'$-$f_{AC}$ measured at nearby temperatures was not observed previously in other functional oxides, such as spinal type ferrites (e.g., $NiFe_2O_4$) [18], hexaferrite (e.g., $Sr_3Co_2Fe_{24}O_{41}$) [19], titanate perovskite (e.g., W-doped $CaBi_4Ti_4O_{15}$) [20], lead zirconate titanate [21], manganese perovskite (e.g., $La_{0.7}Sr_{0.3}MnO_3$ and $La_{0.5}Ca_{0.5-x}Ag_xMnO_3$) [22,23], and $BiFeO_3$ [24].

Accordingly, detecting $R'$-$T$ utilizing a specific $f_{AC}$ within the cross-linking region archives as-proposed $T_{Delta}$- thermistors transport, while reducing or elevating $f_{AC}$ achieves NTCR and PTCR thermistor functionalities [28,29], respectively, as illustrated in Figure 2c. Figure 2d demonstrates the representative $R'$-$T$ tendencies for $SmNiO_3$/$LaAlO_3$ at various $f_{AC}$. For applying low $f_{AC}$ below kHz, a negative $R'$-$T$ tendency, behaving as an NTCR thermistor, with NTCR above 2%/K is observed across the entire target temperature range from 80 K to 300 K. In contrast, applying high $f_{AC}$ exceeding 300 kHz enables the PTCR thermistor functionality, as a positive $R'$-$T$ tendency is observed within the same target temperature range. By imparting an intermediate $f_{AC}$ in between, we obtain a delta-like $R'$-$T$ tendency behaving as a $T_{Delta}$-thermistor, in which case $R'$ firstly increase with $T$ until reaching the maximum at $T_{Delta}$ and afterwards reduces rapidly. The maximum point in $R'$ tendency can be used to lock in the temperature range near $T_{Delta}$, and this is also actively adjustable within the entire temperature range via $f_{AC}$, e.g., increasing $f$ elevates $T_{Delta}$ as more clearly demonstrated in Figure 2e. The above observations are confirmed to be associated to the $SmNiO_3$ film material, rather than the external circuit or the substrate (see confirmation experiments demonstrated in Figure S2 and S3).

It is also worthy to note that the DC or low frequency conductance ($\sigma_{100Hz}$) and the $f_0$ associated to the AC character exhibit a same temperature dependence, as indicated by Figure 2f. The linear fittings of $\ln(\sigma_{100Hz})$-$1000/T$ and $\ln(f_0)$-$1000/T$ indicate the same magnitude and temperature dependence in the activation energy ($E_a$) associated to both DC- and AC- transportations [17-24]. From the quasi-linear $E_a$-$T$ tendency, we further establish differential equation groups to calculate the expressions of Coulomb viscosity for $SmNiO_3$/$LaAlO_3$, as more details demonstrated in the Supporting Information (Section C). A temperature dependence of $\eta_{(T)} \propto T^\beta$ ($\beta=4$) is obtained that indicates an enlarged carrier viscosity within $SmNiO_3$ with the elevated temperature, and this tendency is in consistency with our previous expectation. The pronounced temperature dependence in $E_a$ observed herein in $ReNiO_3$ is in contrast to the ones as previously observed for other functional oxides, e.g., doped ZnO [17], $Sr_3Co_2Fe_{24}O_{41}$ [19] $CaBi_4Ti_4O_{15}$ [20], $La_{0.7}Sr_{0.3}MnO_3$ [22] and $BiFeO_3$ [24].

The insulating phases of other $ReNiO_3$ (e.g. $EuNiO_3$, $Sm_{3/4}Nd_{1/4}NiO_3$ and $HoNiO_3$) exhibit similar functionalities, as their temperature dependent $R'$-$f_{AC}$ and $R''$-$f_{AC}$ curves demonstrated in Figure 3a and Figure S4, respectively. It is worthy to note that reducing the size of rare-earth elements shifts the crosslinking $R'$-$f_{AC}$ region towards a lower frequency, and this is expected to be caused by the strengthening of insulating phase to reduce both the carrier density and viscosity. In addition, similar effect is achieved when imparting biaxial tensile interfacial strain upon the co-lattice grown SmNiO$_3$ on SrTiO$_3$, in which case the insulating phase of $Re$NiO$_3$ can be also strengthened [27,30,31]. For example, the SmNiO$_3$/SrTiO$_3$ sample grown by pulsed laser deposition exhibits same in-plane lattice constants between the film and substrate, as demonstrated by its reciprocal space mapping (RSM) shown in Figure 3b and interfacial morphology shown in the inset of Figure 3c. Such tensile strained SmNiO$_3$/SrTiO$_3$ exhibits a cross-linked region in $R'$-$f_{AC}$ at a lower frequency range, compared to SmNiO$_3$/LaAlO$_3$, as demonstrated in Figure 3c (see respective $R''$-$f_{AC}$ in Figure S6).

In Figure 3d, we summarize the respective $T_{Delta}$-$f_{AC}$ relationships when using SmNiO$_3$/LaAlO$_3$, EuNiO$_3$/LaAlO$_3$, HoNiO$_3$/LaAlO$_3$, Sm$_{3/4}$Nd$_{1/4}$NiO$_3$/LaAlO$_3$ and SmNiO$_3$/SrTiO$_3$ (tensile strained) as $T_{Delta}$-thermistors. From the electronic perspective to achieve active regulations, the $T_{Delta}$ is possible to be elevated via increasing $f_{AC}$ for each individual $Re$NiO$_3$. In addition, from the perspective of materials designs, it is also possible to lower down the required frequency for locking in the same $T_{Delta}$ of $Re$NiO$_3$ by utilizing smaller rare-earth element or establishing biaxial tensile distortions. This enlarges the flexibility to cater for practical electronic applications, e.g. to lock in the same maximum point in the delta-shaped $R$-$T$ tendency (or $T_{Delta}$) via different combinations of the $Re$NiO$_3$ materials and electronically applied $f_{AC}$. Figure 3e demonstrates a representative example for achieving $T_{Delta}$=200 K, e.g. the lowest temperature appears in arctic, for using various $Re$NiO$_3$ materials. The $Re$NiO$_3$/LaAlO$_3$ with a smaller $Re$ requires lower $f_{AC}$ to achieve the same $T_{Delta}$, while as-achieved delta-shape in $R'$-$T$ tendency is observed to be narrower. In contrast, although the tensile strained SmNiO$_3$/SrTiO$_3$ requires lower $f$ compared to SmNiO$_3$/LaAlO$_3$ to achieve the same $T_{Delta}$, its achieved delta-shape in $R'$-$T$ tendency is broader. Utilizing the peak R' in $T_{Delta}$-thermistor we are able to lock in the respective temperature region near $T_{Delta}$ for stabilizing the electronic working circuit or devices.

In conclusion, assisted by temperature dependent NEXAFS we demonstrate that elevating the temperature strengthens the insulating phase of $Re$NiO$_3$ that enhances not only the carrier concentration but also their Coulomb viscosity. This results in cross-linking of $R'$-$f_{AC}$ curves measured at adjacent temperatures and enables the electronic switchability among NTCR-, $T_{Delta}$- and PTCR-thermistor transports by simply regulating $f_{AC}$ without varying the material constitutions or device structure. It is worthy to note that the $T_{Delta}$-thermistor exhibits maxima in $R'$-$T$ tendency that is useful in lock in specific temperatures. From the electronic perspective the $T_{Delta}$ can be further actively elevated via enlarging $f_{AC}$ for a given composition $Re$NiO$_3$ over a broad range of temperature. Furthermore, from the perspective of materials designs, the $T_{Delta}$-$f_{AC}$ relationship as well as the switching frequencies among NTCR-, $T_{Delta}$- and PTCR- thermistors can be widely adjusted via the rare-earth composition and status of interfacial strains. Combining both aspects of AC electronic and materials designs provides large flexibility to cater for intelligent temperature sensing.

**Acknowledgments**
This work was supported by National Natural Science Foundation of China (No. 61674013) and Beijing New-star Plan of Science and Technology (No. Z191100001119071).

**Competing interests**
We declare no competing financial interest.
**Additional information:** Supplementary Information is available for this manuscript.
**Correspondences:** Correspondence should be addressed: Prof. Jikun Chen (jikunchen@ustb.edu.cn).

**Author contributions:** Jikun Chen proposed the original scientific idea, planed for the work, performed partially the experiment (i.e., sample growth, structure and transport characterizations), analysed the data, and write the manuscript; Haifan Li contributed in the transport characterizations under the supervision of Jikun Chen; Jiaou Wang contributed for the TDEXAFS experiment; Xinyou Ke assisted in written of the manuscript and provide constructive suggestions; Jinhao Chen contributed to the sample growth under the supervision of Jikun Chen; Hongliang Dong and Binghui Ge contributed to the TEM experiment; NC and YJ provide supports from the aspects of sample growth and characterization, respectively.

**Availability of data**
The data that support the findings of this study are available from the corresponding author upon reasonable request.

**Figures and captions**

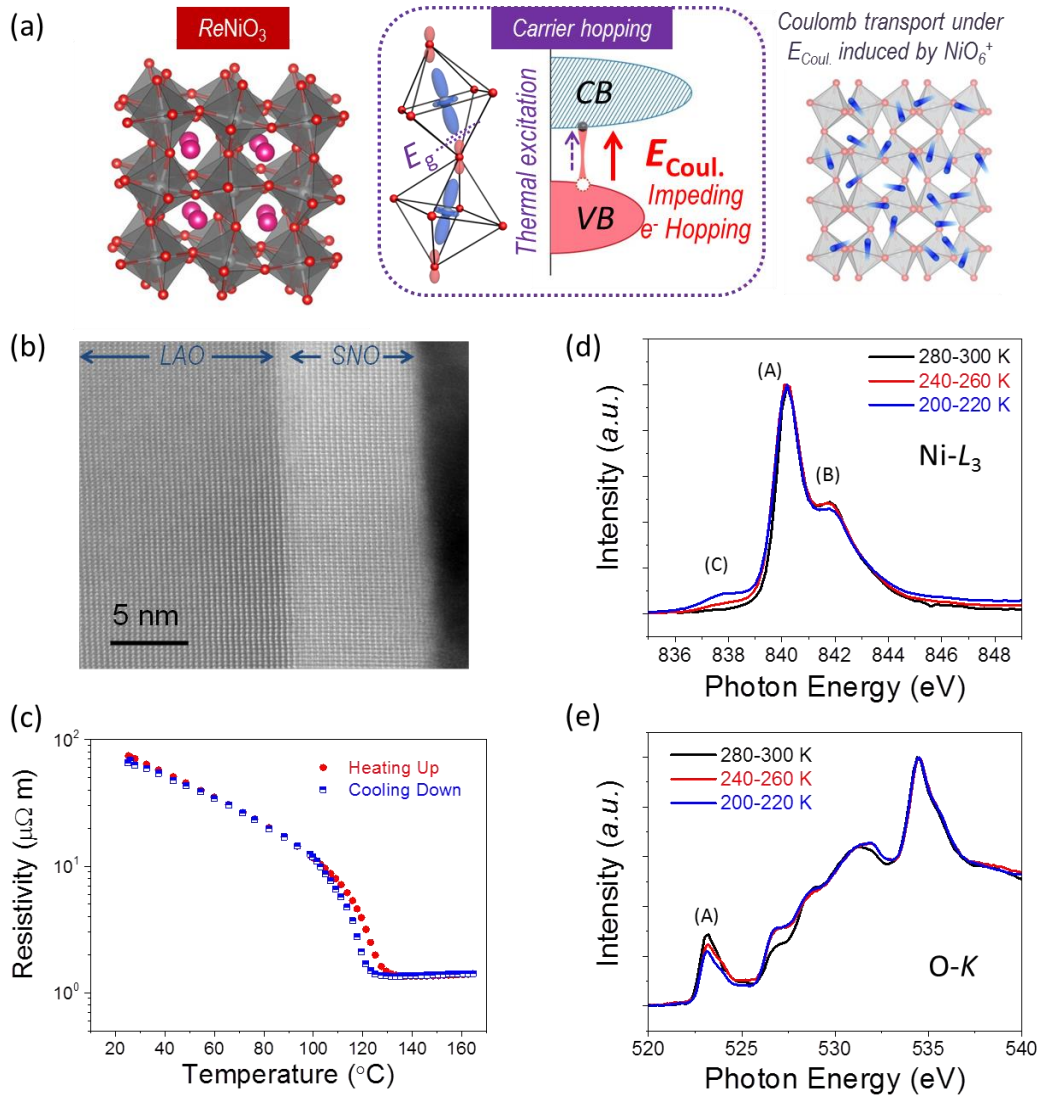

**Figure 1.** **(a)** Schematic illustrations of the crystal structure and lattice Coulomb interaction regulated carrier hopping for *Re*NiO$_3$. **(b)** Interfacial cross-section morphology for as-grown SmNiO$_3$/LaAlO$_3$. **(c)** Temperature dependence of SmNiO$_3$/LaAlO$_3$ measured via both heating up and cooling down process that demonstrates sharp metal to insulator transition was achieved. The inset shows the interfacial morphology for SmNiO$_3$/LaAlO$_3$. **(d),(e)** Temperature dependent near edge X-ray absorption fine structure analysis of **(d)** Ni-$L_3$ edge and **(e)** O-$K$ edge of SmNiO$_3$ at various temperatures.

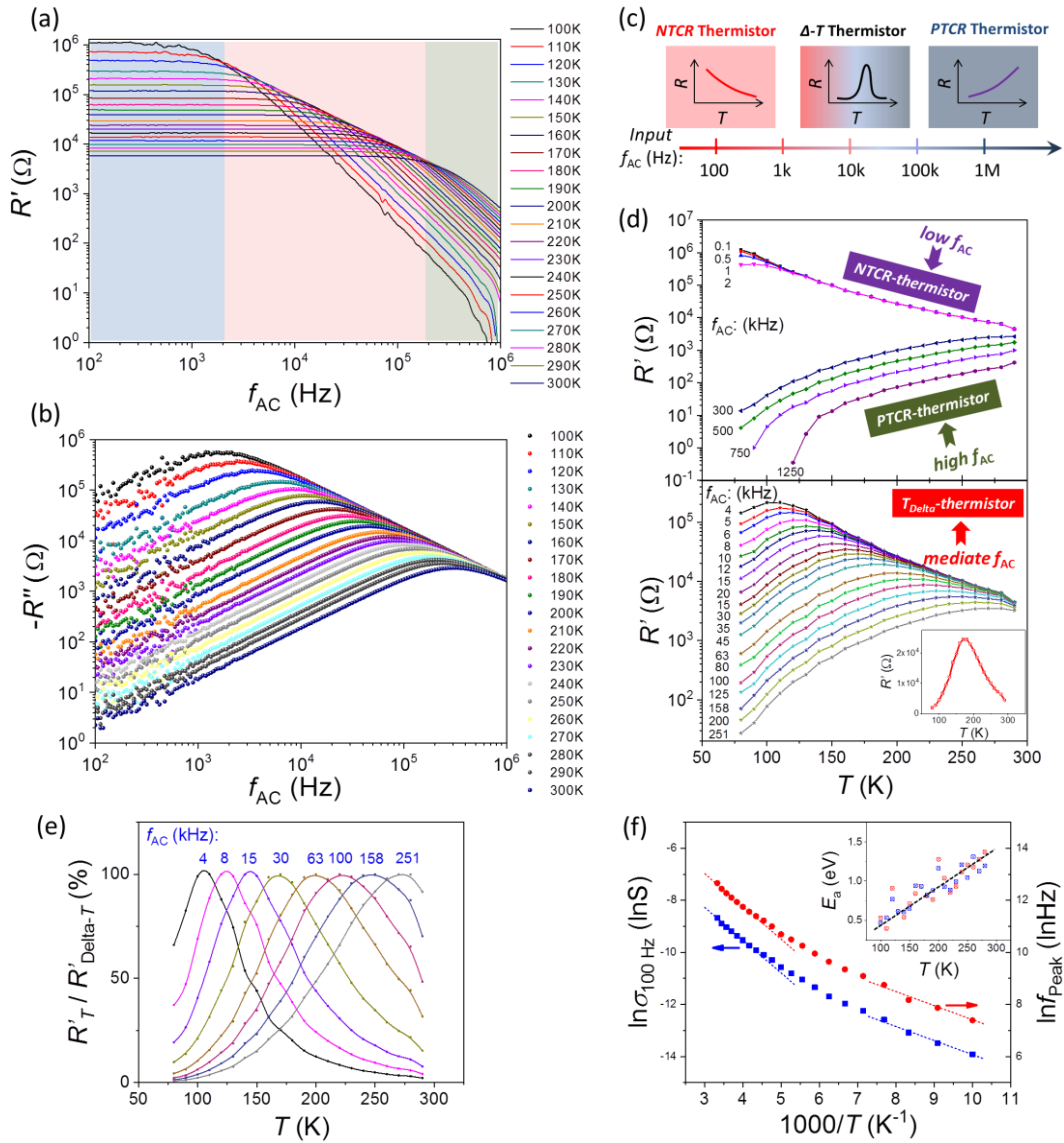

**Figure 2.** (**a**),(**b**) The (**a**) real and (**b**) imagination part of the impedance measured for SmNiO$_3$/LaAlO$_3$ as a function of frequency of the input detection AC-signal at various temperatures. (**c**) Illustrating the AC-frequency determination and switchability among NTCR, TDelta- and PTCR thermistor functionalities. (**d**) Temperature dependence in resistance ($R'$-$T$) measured at various AC-frequencies. (**e**) Relative resistance compared to their maximum magnitudes at $T_{\text{Delta}}$ for applying various AC-frequencies. (**f**) The ln($\sigma_{100\text{Hz}}$)-1000/$T$ and ln($f_0$)-1000/$T$ tendencies that are used to calculate the temperature dependence in the activation energy ($E_a$) from both perspective of the DC and AC transportations.

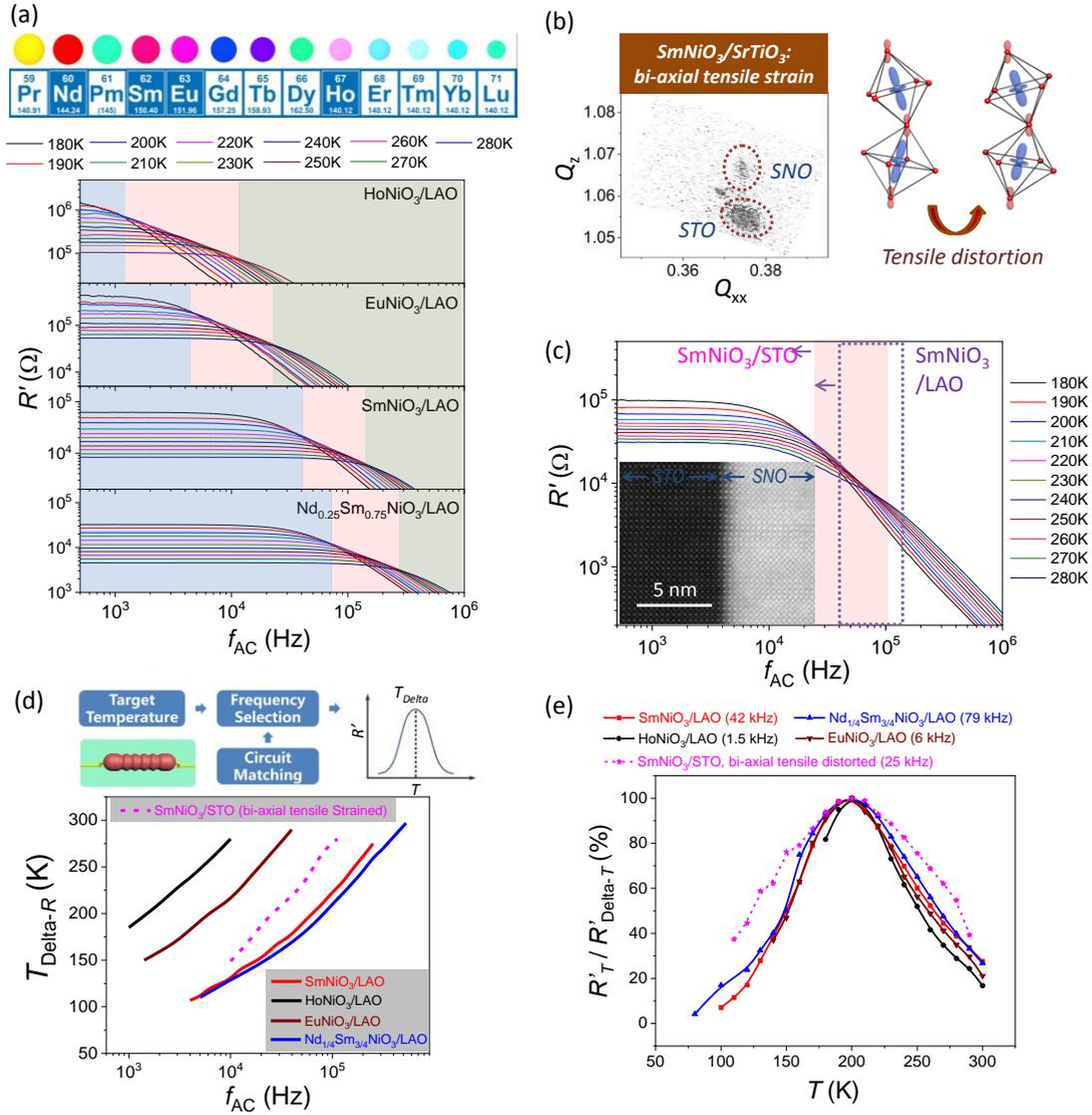

**Figure 3.** (a) Comparing the resistance to AC frequency ($R'$- $f_{AC}$) tendency measured at various temperatures for $Re$NiO$_3$/LaAlO$_3$ with various rare-earth compositions including Nd$_{1/4}$Sm$_{1/3}$, Sm, Eu, Ho. (b) Reciprocal apace mapping (RSM) for the pulsed laser deposited SmNiO$_3$ coherently on SrTiO$_3$ (001) substrate, in which case biaxial tensile distortion is imparted on as-grown thin film. (c) Comparing the resistance to AC frequency ($R'$- $f_{AC}$) tendency measured at various temperatures for the biaxial tensile strained SmNiO$_3$/SrTiO$_3$ to the one for SmNiO$_3$/LaAlO$_3$ at slight biaxial compressive distortion. The inset in (c) demonstrate a coherent interface achieved for SmNiO$_3$/SrTiO$_3$ that further confirms the preservation of tensile interfacial strain. (d) Summarizing the required AC-frequency to achieve the desired $T_{Delta}$ when using $Re$NiO$_3$ thin films at various rare-earth compositions and states of interfacial strains as a $T_{Delta}$ thermistor. The upper figure illustrates the principle to apply the $T_{Delta}$-thermistor in electronic circuit. (e) A representative case for locking 200 K (the lowest temperature in arctic) utilizing the $T_{Delta}$ thermistor of various $Re$NiO$_3$ thin films.